\documentclass[12pt]{article}

\global\arraycolsep=1pt
\def\fnote#1#2{\begingroup\def\thefootnote{#1}\footnote{#2}\addtocounter
{footnote}{-1}\endgroup}

\usepackage{amsbsy,amssymb,latexsym}

\newcommand{\beq}{\begin{equation}}
\newcommand{\eeq}{\end{equation}}
\newcommand{\beqa}{\begin{eqnarray}}
\newcommand{\eeqa}{\end{eqnarray}}
\newcommand{\bR}{{\mathbb R}}
\newcommand{\bZ}{{\mathbb Z}}

\newcommand{\CP}{{\mathbb C}P}

\def\fnote#1#2{\begingroup\def\thefootnote{#1}\footnote{#2}\addtocounter
{footnote}{-1}\endgroup}

\usepackage{amsbsy,amssymb,latexsym}

\begin{document}
\begin{flushright}
OCU-PHYS 215 \\
hep-th/0407114\\

\end{flushright}
\vspace{15mm}

\begin{center}
{\bf\Large
Sasaki-Einstein Twist
of
Kerr-AdS Black Holes
}

\vspace{20mm}
Yoshitake Hashimoto\fnote{$\dagger$}{
\texttt{hashimot@sci.osaka-cu.ac.jp}
},
Makoto Sakaguchi\fnote{$\star$}{
\texttt{msakaguc@sci.osaka-cu.ac.jp}
}
and
Yukinori Yasui\fnote{$\ast$}{
\texttt{yasui@sci.osaka-cu.ac.jp}
}
\vspace{10mm}

\textit{
${{}^\dagger}$
Department of Mathematics, Osaka City University
}
\vspace{2mm}

\textit{
${{}^\star}$
Osaka City University
Advanced Mathematical Institute (OCAMI)
}
\vspace{2mm}

${}^\ast$
\textit{
Department of Physics, Osaka City University
}

\vspace{5mm}

\textit{
Sumiyoshi,
Osaka 558-8585, JAPAN}
\end{center}
\vspace{15mm}

\begin{abstract}
We consider Kerr-AdS black holes
with equal angular momenta
in arbitrary odd spacetime dimensions
$\ge 5$.
Twisting
the
Killing vector fields of the 
 black holes,
we reproduce the compact
Sasaki-Einstein manifolds
constructed by Gauntlett, Martelli, Sparks and Waldram.
We also discuss an implication of the twist
in string theory and M-theory.
\end{abstract}
\newpage

Kerr-AdS black holes
are
characterized by mass,
angular momenta and cosmological constant.
In spacetime dimension $d$,
the number of angular
momenta is equal to the rank of the rotation group SO$(d-1)$.
The five-dimensional
Kerr-AdS black holes with two angular momenta were
constructed
in \cite{HHTR},
and recently the general form in arbitrary dimension was found by
using the Kerr-Schild ansatz \cite{GLPP}.

On the other hand, the Wick rotation of the black holes leads to
Riemannian
metrics. However, the metrics in general
do not extend smoothly to compact manifolds.
In \cite{Page}\cite{HSY}\cite{GLPP}, it was shown that this can be achieved by
taking a certain limit (Page limit) which enhances the isometry
of the metric. Indeed, the infinite series of 
Einstein metrics on 
compact manifolds were explicitly constructed \cite{HSY}\cite{GLPP},
and
analyzed in detail in \cite{GHY}.

Recently, infinite series of
Sasaki-Einstein metrics on compact manifolds
were presented in \cite{GMPW:Sasaki}\cite{GMPW:A new}.
It is expected that these metrics can be related to
some Kerr-AdS black holes by a certain limit.
Our aim in this letter is to clarify the relation
between them.


We begin with the
$(2n+3)$-dimensional Kerr-AdS
black hole with a negative
cosmological
constant $(2n+2)\lambda <0$ ($n\ge 1$) as follows~\cite{HHTR}\cite{GLPP}:
\begin{eqnarray}
\hat g=-\frac{\hat W(r)}{\hat b(r)} dt^2 +
\frac{dr^2}{\hat W(r)}
+r^2\left(
g_{\CP^n}
+\hat b(r)
\left(
d\psi+A + \hat f(r)dt
\right)^2\right),
\label{BH}
\end{eqnarray}
where 
\begin{eqnarray}
\hat W(r) &=&
1- \lambda r^2 -\frac{2 M(\delta^2 + \lambda J^2)}{r^{2n}}
+\frac{2 M J^2}{r^{2n+2}}
=(1-\lambda r^2)\hat b(r)-\frac{2M\delta^2}{r^{2n}}
,\nonumber\\
\hat b(r) &=&
1+\frac{2  MJ^2}{r^{2n+2}},\nonumber\\
\hat f(r) &=&
\frac{1}{J}\left(1-\frac{\delta}{\hat b(r)} \right).
\end{eqnarray}
The metric $g_{\CP^n}$ is the Fubini-Study metric on
$\CP^n$ with a normalization $Ric_{\CP^n}=(2 n+2)g_{\CP^n}$,
and the 1-form $A$ is the U(1) connection
associated with the K\"ahler form $dA/2$ 
on  $g_{\CP^n}$.
The black hole is parameterized by the mass $M$,
the angular momentum $J$ and a trivial parameter
$\delta$.
The parameter $\delta$
is related to the parameter $\beta$
introduced in  \cite{CLP}
as $ \delta= -\lambda J^2 \beta +1$.
This metric is a special case that 
all angular momenta are set to be equal.

The metric (\ref{BH})
reduces to the AdS metric at $r\to\infty$
because the metric of the circle bundle over $\CP^n$
tends to the standard metric of $S^{2n+1}$.\fnote{$\natural$}{
If we replace the Fubini-Study $\CP^n$
by an
arbitrary Einstein-K\"ahler manifold
with the same scalar curvature,
we obtain another Kerr black hole with different asymptotic
behavior.}
A horizon appears for sufficiently small $J$.
If
we set $\delta^2=-\lambda J^2$,
the $\hat W(r)$ does not have positive roots
so that
the curvature singularity at $r=0$
is not screened by the horizon, and so is naked.
As will be seen below, in the Euclidean picture
this solution is shown to be related to the Sasaki-Einstein metrics.

The Euclidean Einstein  metric
 with a positive cosmological
constant $(2n+2)\lambda >0$
is extracted from the
Kerr-AdS black hole (\ref{BH}) by the substitution
$t \rightarrow i \tau$ and  
$J \rightarrow i J$:
\begin{eqnarray}
g=\frac{W(r)}{b(r)} d\tau^2 +
\frac{dr^2}{ W(r)}
+r^2
\Bigl(
g_{\CP^n}
+b(r)
\left(
d\psi+A + f(r)d\tau
\right)^2
\Bigr),
\end{eqnarray}
where 
\begin{eqnarray}
W(r) &=&
1- \lambda r^2 -\frac{2 M(\delta^2 - \lambda J^2)}{r^{2n}}
-\frac{2 M J^2}{r^{2n+2}}
=(1-\lambda r^2) b(r)-\frac{2M\delta^2}{r^{2n}}
,\nonumber\\
b(r) &=&
1-\frac{2  MJ^2}{r^{2n+2}},\nonumber\\
f(r) &=&
\frac{1}{J}\left(1-\frac{\delta}{b(r)} \right).
\end{eqnarray}
The metric has the isometry SU($n+1)\times$U(1)$\times \bR$.
The generator of U(1)$\times \bR$ is
given by
$(\frac{\partial}{\partial \psi},\frac{\partial}{\partial \tau})$. 
It is easy to see that under the Page limit
and a special choice of the parameters \cite{Page}\cite{HSY}\cite{GLPP}
this metric reduces to a homogeneous Einstein 
metric with the isometry SU($n+1$)$\times$SU($2$)$\times$U(1)
on a  circle bundle
over $\CP^n\times S^2$.
Indeed, the metric can be written as
\begin{eqnarray}
g_0&=&
\frac{1}{W_0}(d\chi^2+\sin^2 \chi d\eta^2)
+r_0^2\left(
g_{\CP^n}
+b_0(d\psi+A+\frac{k}{2}\cos\chi d\eta)^2
\right),
\end{eqnarray}
where
\begin{eqnarray}
W_0&=&\frac{2\lambda (n+1)\left( 2(n+1)-(n+2)b_0\right)}{n+1-b_0},
\nonumber\\
r_0^2&=&
\frac{n+1-b_0}{\lambda (n+1)},\nonumber\\
k&=&\pm\frac{2\sqrt{(n+1)b_0(1-b_0)}}{b_0(2(n+1)-(n+2)b_0)}
\in \bZ~,
\end{eqnarray}
and $b_0$ is a constant with $0<b_0<1$.
In the case of $n=1$, this reproduces the metric given in
Theorem 2 of \cite{HSY}.
Further, for $k=1$,
it gives the homogeneous Sasaki-Einstein manifold $T^{1,1}$.
 
We now transform the metric to inhomogeneous 
Sasaki-Einstein metrics
on circle bundles over $\CP^n \tilde{\times} S^2$ ($S^2$ bundle
over $\CP^n$)
presented in \cite{GMPW:Sasaki}\cite{GMPW:A new}.

First, we set $\delta^2=\lambda J^2$,
then the 
coefficient of $1/r^{2n}$ in $W$ vanishes.

Twisting
the U(1)$\times \bR$ coordinates as
\beq
\tilde \tau={\tau} +  J \psi,
\eeq
we obtain
\beq
g=g_{K}
+(J d\psi -\sigma)^2,
\label{Sasaki:0}
\eeq
where the metric $g_{K}$ is a local positive K\"ahler-Einstein
metric in dimension $2n+2$,
\beq
g_{K}=\frac{dr^2}{W(r)}+ r^2 g_{\CP^n} +
r^2 W(r)\left(\frac{d\tilde{\tau}}{J} + A\right)^2,
\eeq
and the K\"ahler form of $g_K$ is given by
$d \sigma/2 \sqrt{\lambda}$,
\beq
\sigma= \left( 1 -\frac{\sqrt{\lambda} r^2}{J} \right)d\tilde\tau
-\sqrt{\lambda}r^2 A.
\eeq
Thus, as is well known,
 the metric $g$ in (\ref{Sasaki:0}) turns out to be
locally Sasaki-Einstein.
If we write the metric $g$ by the coordinates $(\tau, \tilde\tau)$,
instead of $(\tau,\psi)$ or $(\tilde\tau, \psi)$,
we can eliminate the parameter $\delta$ after rescaling 
$M\delta^2\to M$ and $J\delta^{-1}\to J$~\fnote{$\ddagger $}{
The authors are grateful to Gary Gibbons, Malcolm Perry and Chris Pope
for this remark.
}.

On the other hand, twisting the coordinates as
\beq
\tilde \psi={\psi}-\frac{c}{J}\tau,
\eeq
we have
\beq
g=g_{C}+\omega(r)\left(
d \tau+f(r)( d\tilde{\psi}+A)
\right)^2,
\label{Sasaki}
\eeq
where
\beq
g_{C}=\frac{dr^2}{W(r)}+r^2 g_{\CP^n}+
q(r)( d\tilde{\psi}+A)^2,
\eeq
and the components are given by
\beqa
\omega(r)&=& k^2 r^2 W(r)+
(k \sqrt{\lambda}r^2-1)^2,\nonumber \\
f(r) &=& \frac{r^2}{\omega(r)} 
\left( k W(r)+\sqrt{\lambda}
(k \sqrt{\lambda}r^2-1) \right), \nonumber \\
q(r) &=& \frac{r^2 W(r)}{\omega(r)}
\eeqa 
with $k=(c+1)/J$.
The metric $g_C$ is conformally K\"ahler \cite{GMPW:A new}.

The singularities coming from
the roots $r=r_{i}$ of $W=0$
can be resolved by the restriction of the range of the angle
$\tilde\psi$;
putting $R^2=4(r-r_i)/W'(r_i)$ one has in the limit
$r \rightarrow r_i$,
\beq
\frac{dr^2}{W(r)^2}+q(r) d\tilde{\psi}^2 \rightarrow
dR^2 + K_i^2 R^2 d\tilde\psi^2,
\eeq
where
\beq
K_i=\frac{(n+2)\lambda r_i^2-(n+1)}
{k \sqrt{\lambda}r_i^2-1}.
\eeq
If we set $\lambda(n+2)/(n+1)=k \sqrt{\lambda}$,
that is,
\begin{eqnarray}
c=\frac{n+2}{n+1}\sqrt{\lambda}J-1,
\end{eqnarray}
 then $K_i$ is independent of $r_i$.
Under
a suitable condition on the parameter $MJ^2$, 
the corresponding metric $g$ has an SU($n+1$)$\times$U(1)$\times$U(1)
symmetry, and it
 reproduces a Sasaki-Einstein metric on a compact manifold
given by  Gauntlett et al. in \cite{GMPW:Sasaki}\cite{GMPW:A new}.

\smallskip

We shall comment on the implication of our method in the
higher dimensional context.
As explained above, the higher dimensional backgrounds
are related each other as follows:

\begin{minipage}{70mm}
\begin{eqnarray*}
\mbox{AdS}_p&\times& S^q\\
&\updownarrow & \mbox{Wick rot.}\\
H^p&\times&\mbox{dS}_q\\
&\updownarrow & \mbox{cosmo.}\\
S^p&\times&\mbox{AdS}_q
\end{eqnarray*}
\end{minipage}
~~~
\begin{minipage}{70mm}
\begin{eqnarray*}
\mbox{AdS}_p&\times& M_{SE}^q\\
&\uparrow & \mbox{Wick rot. and $\delta^2=\lambda J^2$}\\
H^p&\times&M_{dS}^q\\
&\updownarrow & \mbox{cosmo.}\\
S^p&\times&M_{AdS}^q
\end{eqnarray*}
\end{minipage}
\\

\noindent
where $(p,q)=(5,5)$, $(4,7)$,
and a $p$-form flux is associated with them.
The left hand side shows that
the maximally supersymmetric backgrounds
are related to each other.
Under the Wick rotation and a sign change
of the cosmological constant,
the AdS$_5\times S^5$ solution
in the type-IIB string theory
is mapped to itself,
while AdS$_4\times S^7$ becomes
 $S^4 \times$AdS$_7$.
In the right hand side,
we have generalized  $S^q$ to  $M^q_{SE}$,
where $M_{SE}^q$
stands for the $q$-dimensional Sasaki-Einstein manifold
specified by (\ref{Sasaki}).
This shows the relation
between  AdS$_{p}\times M_{SE}^q$
and  $S^{p}\times M_{AdS}^q$,
where $M_{AdS}^q$
means  the $q$-dimensional Kerr-AdS black hole.
It is known that the former solution
admits supersymmetry
due to the Sasaki-Einstein structure of $M_{SE}^q$,
and that string/M-theory
on  AdS$_{p}\times M_{SE}^q$
is  dual
to supersymmetric Yang-Mills theory in $(p-1)$-dimensions.
Though the condition $\delta^2=\lambda J^2$
implies a naked singularity for $M_{AdS}^q$
and cannot be imposed consistently on  $M_{dS}^q$,
where  $M_{dS}^q$ means the Kerr-dS black hole,
it may be interesting to examine
string/M-theory on  $S^{p}\times M_{AdS}^q$
and the dual Yang-Mills theory.

\vspace{4mm}

MS gave a talk in ``Quantum Field Theory 2004"
(July 13-16)
at Yukawa Institute
for Theoretical Physics.
The authors thank the organizers
for giving the chance
and participants for useful comments.
This paper is supported by the 21 COE program
``Constitution of wide-angle mathematical basis focused on knots".
Research of Y.H. is supported  in part by the Grant-in
Aid for scientific Research (No.~14540088, No.15540040 and No.~15540090)
from Japan Ministry of Education.
Research of Y.Y. is supported  in part by the Grant-in
Aid for scientific Research (No.~14540073 and No.~14540275)
from Japan Ministry of Education.


\end{document}